\documentclass[aps,pre,onecolumn,showpacs,showkeys,superscriptaddress,endfloats,floatfix]{revtex4}
\usepackage{amsmath, amsthm, amssymb, amsfonts}
\usepackage{mathrsfs}
\usepackage{graphicx}
\usepackage{bm}

\newcommand{\beq}{\begin{equation}}
\newcommand{\eeq}{\end{equation}}

\begin{document}

\title{Correlated flares in models of a magnetized ``canopy''}

\author{Marco Baiesi}
\affiliation{Dipartimento di Fisica, Universit\`a di Firenze,
I-50019 Sesto Fiorentino, Italy}
\affiliation{INFN, Sezione di Firenze,
I-50019 Sesto Fiorentino, Italy}

\author{Christian Maes}
\affiliation{Instituut voor Theoretische Fysica, K.U.Leuven, Belgium}

\author{Bidzina M. Shergelashvili\footnote{On
leave from Georgian National Astrophysical
Observatory, Kazbegi ave. 2a, 0160 Tbilisi, Georgia}}
\affiliation{Instituut voor Theoretische Fysica, K.U.Leuven, Belgium}

\begin{abstract}
A model of the Lu-Hamilton kind is applied to the study of
critical behavior of the magnetized solar atmosphere. The main
novelty is that its driving is done via sources undergoing a
diffusion. This mimics the effect of a virtual turbulent substrate
forcing the system. The system exhibits power-law statistics not
only in the size of the flares, but also in the distribution of
the waiting times.
\end{abstract}

\maketitle

\section{Introduction}

One of the most interesting properties of spatially extended
dynamical systems in nature is that they can exhibit critical
behavior.  That terminology loosely derives from the theory of
phase transitions in equilibrium statistical mechanics but has
come to mean more generally that the system manifests power-law
statistics in its characteristic space-time distributions.  There
may be long range or long term correlations and the structure of
fluctuations could have very nonlocal features.

  There is so far
no general theory dealing with critical behavior for general
nonequilibrium systems.  At this moment certain schemes are tried,
even for some aspects seemingly unrealistic ones, and it is hoped
that they retain some general validity when confronted with a
larger circle of phenomena. One of the attempts for describing in
a more unified way nonequilibrium and dynamical power-laws has
been called self-organized criticality
(SOC)~\cite{BTW1987,bak,sornette,jenbook}.
Normally SOC may be
expected in slowly loaded extended systems with local
instabilities evolving according to a threshold dynamics, namely
being active only when some level of ``stress'' is larger than a
threshold. Local instabilities may trigger further ones upon
relaxing, generating avalanches of relaxation that bring the
system from a metastable state to another. The occurrence of
scale-free avalanches has been well documented in cellular
automata, often named sandpiles, which provide a  surprisingly
simple way of simulating SOC
\cite{BTW1987,bak,sornette,jenbook}. The scale-invariant
distribution of avalanches in sandpiles is the hallmark of SOC,
and it is a robust feature, suggesting that avalanching processes
can be valid explanations of several natural scale-free
phenomena~\cite{bak}. For example, one can argue that the
power-law distribution of the  energies released by earthquakes or
by solar emissions are due to the avalanching nature of these
processes.

The present paper revisits some earlier attempts of modeling the
critical state of the solar atmosphere via SOC models. The
atmosphere of the Sun is very complex and inhomogeneous. Even
though there are a growing amount of data concerning solar flare
activity, e.g. in \cite{aschwanden1,aschwanden2}, we still lack
detailed information about statistical-topological aspects. The
spatial and temporal resolution of the observation are too
``rough'' for the detection of the small scale structures of the
solar atmosphere participating in the considered processes at
either photospheric, chromospheric or coronal levels. Various
first questions have not been answered. For example, the manifest
dynamical features of the solar activity or the mechanisms of
heating of the outer atmosphere have not been resolved to a
sufficient degree.  However, it is believed that the heating and
the eruptive phenomena in the solar atmosphere are
related to magnetic structures that are constantly being driven
and that dissipate via reconnection and wave mechanisms
\cite{heyvaerts2001} (see also recent new developments reported in
\cite{sherg1,sherg2}). While simplified models must obviously be
treated with caution, they can also be welcomed as
  highlighting single essential features.

The relevance of SOC models for the study of the solar atmosphere
has been realized since the pioneering work of Lu and Hamilton
\cite{LH1991} (see also~\cite{LU1993}). We will
refer to it as the LH-model.  The idea was to develop a
cellular automaton model for the solar atmosphere that would
realize some of the heuristics and of the ideas stated (i) in
\cite{parker1989,sturrock1984} that solar flares
might represent a cascade of smaller events of magnetic
reconnection and (ii) in \cite{parker1988,sturrock1990}
that in the coronal heating a big number of
small non-thermal events could make a significant contribution.
These works initiated investigating whether cascades of small size
dissipations of the magnetic field can avalanche in solar flares
to support the observed dynamics and heating rate of the solar
atmosphere. The LH-model indicated that under certain conditions
for a 3D domain that is slowly ``fed'' by the magnetic field,
the system evolves into a critical state showing power-law
statistics in the energy released by avalanches (flares).

 While these first attempts in the context of the solar
atmosphere had opened the possibility  to model SOC events under
coronal conditions, there were also several and significant
limitations. As was pointed out in
\cite{isliker2000,vlahos2004,podlad2002,krasnosels2002}, the
LH-model faced some difficulties. For example,
there was a problem with the
correct physical interpretation of the applied magnetic field. On
the other hand the latter authors suggested to consider a large 2D
domain, which is uniformly fed by sources of different type and
topology.
Moreover, it has been emphasized ever since
\cite{boffetta1999} that the LH-model and other sandpile models
have time series with exponentially separated events. Hence they
do not reproduce real waiting time probability distributions.
That obviously has casted doubts on whether the concept and
modeling of SOC is useful at all for studying the dynamical
processes in the solar atmosphere.

Recently in \cite{BM2006} it has been suggested that the basic
reason for the unrealistic temporal statistics of some sandpiles
comes from the feeding uniformly randomized in space and time.
Indeed, this feature is likely to be artificial in several
contexts. For example, earthquake epicenters are clustered in
space and time. Thus, it appears natural that SOC models will not
display clustering and correlation of events in time when a
randomization in space of the ``epicenters'' is forced by the
chosen driving. A new sandpile cellular automaton was devised
having a more natural feeding mechanism, namely a feeding
associated to the position of a random walker, mimicking the
spatial correlations of diffusing epicenters. The result was a
time series with correlated avalanches~\cite{note_1}, in
particular with power-law tails in the waiting time distributions,
which also collapse onto a single scaling function when rescaled
by the rates of events, as found for
earthquakes~\cite{earthquakes1,earthquakes2} and solar
flares~\cite{flares}.

Existing SOC models imply a rather simplified configuration of the
magnetic fields and of the external drivers supporting the system.
Moreover the updating mechanisms are only qualitatively
representing some of the very complex magnetic dynamics.  While we
continue in the same SOC-tradition of mathematical modeling, we
add here the novel feature discussed above, namely the diffusing
feeding sources. Thus, they
are not fixed in space nor do they jump from one
site to another in an absolutely random way, but they perform a random
walk motion, which is the prototype of a correlated evolution
in space and time.
In the present work we consider a sandpile model of a local area of the solar
atmosphere in which we have a two-dimensional slice through which
the perpendicular components of  the magnetic field lines
contribute to the dynamics.  The feeding should be thought of as
the complicated result of a turbulent substrate. Therefore, a
diffusive feeding seems suited for this problem as well, giving an
additional temporal scale in the system, the rate of source
diffusion.

The work is organized as follows: in the next section we present
the details of our models and the specifics of the simulation
code. In Section \ref{sec:result} we discuss the results of our
simulations, after which we conclude.

\section{The model}

Our model is similar to the one considered by Lu and Hamilton~\cite{LH1991}.
The differences are as follows: (i) we take a
two-dimensional square lattice of side $L$. (ii) The feeding is
not spatially uniform but its position is subject to a random walk.
(iii) Additionally, after studying the standard case in which the boundary
conditions are open and one perturbs the system with a source of given sign,
we also investigate the case in which two sources of opposite polarity
perturb the system.

The magnetic field $h(i)$ at site $i$ is thought to be
orthogonal to the two-dimensional domain.
We consider each site in combination with its $z=4$ neighboring
ones.  For a lattice model, the Laplacian
of the magnetic field is defined as
\beq \label{lgrad}
\Delta h(i) = \frac{1}{z+1}\sum_{|j-i|=1} [ h(j) - h(i) ].
\eeq
This quantifies the local curvature of the field at site $i$
and if it is too high,
\beq
|\Delta h(i)| > C
\label{eq:inst}
\eeq
(we set the energy scale by choosing $C=1)$, by definition
there is an instability in the local magnetic field.
Thus, this approach attempts to model, within a restricted set-up,
the dissipation of high electric currents. We do not include
e.g.~the reconnection mechanism, which would require a more
sophisticated modeling.

Instabilities arise in the system because of some feeding mechanism.
In a first model (model~I),
a perturbation chosen in the range $[0,\delta]$ is added to a site  $i_+$
at each time step.
This corresponds to the standard LH perturbation. As in the LH-model,
we use open boundary conditions.
On the other hand, in ``model~II'' we also allow
another source to put independently
a perturbation chosen in the symmetric range $[-\delta,0]$ at a site $i_-$.
As a result, in this case the average field is zero.
The double feeding is done simultaneously at each time step.
Furthermore, in model~II we use periodic boundary conditions.

An instability (\ref{eq:inst}) is resolved by setting $h(i)$
and all its neighbors $h(j)$ equal to the local average field,
\begin{align}
h(i)  \to & \;\overline{h}(i) & \nonumber\\
h(j)  \to & \;\overline{h}(i) & \text{for all $j$ neighbor of $i$}
\label{eq:relax}
\end{align}
with
\beq
\overline{h}(i) = \frac{1}{z+1} \left[ h(i) +
\sum_{|j-i|=1} h(j) \right] = h(i) + \Delta h(i) \;.
\label{eq:meanh}
\eeq
This redistribution can in turn induce instabilities to neighboring sites,
eventually generating avalanches of updates. We chose to do the transitions
(\ref{eq:relax}) in parallel for all sites $i$ resulting
unstable, as in the LH-model,
iterating this process until the relaxation is over
(when all sites are stable again).
The whole set of updates constitutes an avalanche at a given time step, and
the total number local relaxations (\ref{eq:relax})
is the size of the avalanche.
In the next time step, with probability $p_{move}$,
$i_+$ (and independently $i_-$, in model II)
jump to a nearest neighbor within the $L\times L$ square.
Note that the LH-model instead would pick at random a new $i_+$ from
all the sites of the lattice.

 In real observations, it is not very
simple to define clearly the time and the amplitude of a solar
flare. For our model that more or less amounts to making a
reasonable convention. As a measure of the strength of an
avalanche, we use the size, but we checked that on average
it scales linearly with the released magnetic energy,
which is the sum of all releases of magnetic energy by
local relaxations of an avalanche.

In order to define waiting times between avalanches, it is
important first to fix the scale of events to be
studied~\cite{PACZUSKI2005}. One can say that events with size
$<s_{min}$ compose quiet periods where no avalanches
are``detected'', and waiting times $t_w$ between avalanches can be
defined as the number of seed additions between two avalanches.
(One should write $t_w^{s_{min}}$ with an index recalling that a
threshold $s_{min}$ is used, but it is omitted for simplicity).
The introduction of many thresholds $s_{min}$ allows for more
detailed analysis of the time series of events, eventually leading
to the discovery of scale-invariant
properties~\cite{earthquakes2,earthquakes1,flares}.

\section{Results}
\label{sec:result}

We have two characteristic scales of the driving mechanism. The
first one sets a typical scale of the perturbation, as given by
$\delta$.  In this study we set $\delta = 1$. The
other scale is diffusive and comes with the mobility $p_{move}$
(diffusion constant) of the feeding sources. Our results will depend on
this parameter.

\subsection{Model I}

We first discuss what we obtain with model~I. It is not our purpose to
show a comprehensive spectrum of results for many choices of the parameters.
We just present the
results that we believe are the most interesting for our discussion.

\begin{figure}[htb]
\includegraphics[angle=0,width=7.8cm]{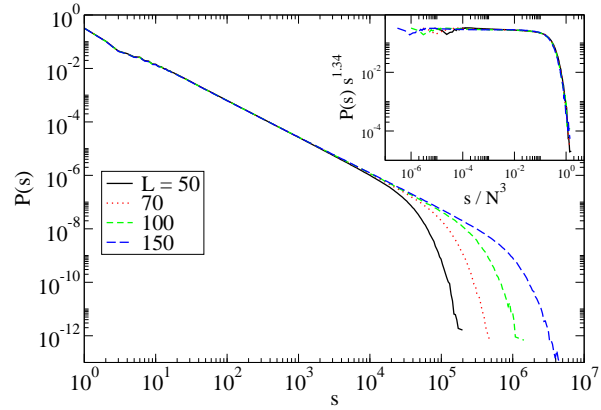}
\caption{Log-log plot of the probability distributions of
avalanche sizes for model~I, for various $L$'s and for $p_{move} =
10^{-1}$. Inset: data collapse of distributions rescaled according
to (\ref{eq:fss}). \label{fig:M1_Ps} }
\end{figure}

First in Fig.~\ref{fig:M1_Ps}
we show the size distributions for $p_{move}=10^{-1}$ and several $L$'s,
$P_L(s)$.
These distributions are power-laws cutoff at a size $\sim L^\nu$ with
$\nu=3$, and the
good data collapse in the inset of Fig.~\ref{fig:M1_Ps} shows that an
exponent $\gamma \simeq 1.34$ can be used to describe the size distribution
with a scaling form
\beq
P_L(s)\sim s^{-\gamma} F(s/L^\nu)
\label{eq:fss}
\eeq
that is typical of SOC systems.
In particular, one extrapolates a diverging power-law range for $L\to\infty$.

\begin{figure}[htb]
\includegraphics[angle=0,width=7.8cm]{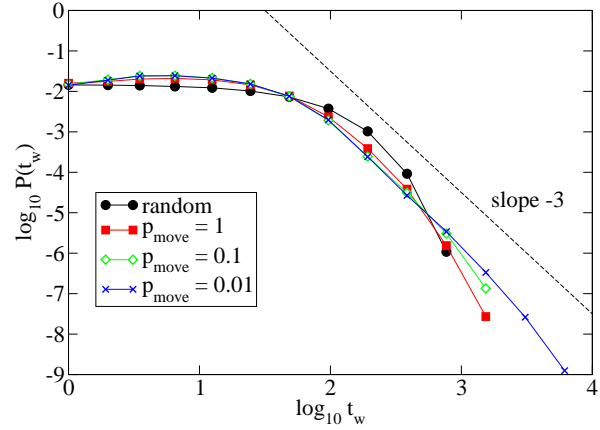}
\caption{Log-log plot of the probability distributions
of waiting times for model~I with $L=100$, threshold $s_{min}=3000$,
and for several $p_{move}$.
Square symbols refer to the same distribution for the
model with feeding randomly distributed in space.
The straight dashed line represents a power-law $\sim t_w^{-3}$.
\label{fig:M1_Ptw} }
\end{figure}

In Fig.~\ref{fig:M1_Ptw} we show the distributions of waiting times,
$P(t_w)$,
between event with size larger than $s_{min}=3000$, for a lattice
with side $L=100$, and three
values of the source moving rate $p_{move}=1$ (filled squares
connected with dotted lines), $p_{move}=10^{-1}$ (empty squares
connected with dotted lines), $p_{move}=10^{-2}$ (crosses
connected with dotted lines).
For comparison we also plot the distributions corresponding
to  random uniform feeding, which are almost exponential, as
previously found~\cite{boffetta1999}.
The other curves instead exhibit a power-law tail
that becomes wider for decreasing $p_{move}$.
This behavior could be expected, as a small $p_{move}$ implies a slower
diffusion of the sources and thus a higher spatial correlation between the
sites were avalanches occur.


In Fig.~\ref{fig:M1_Ptw_s}  we see the same distribution functions
corresponding to a fixed value of the source moving rate $p_{move}=10^{-1}$
and different thresholds. The simulations are done for
the lattice with size $L=150$. The absolute value of the exponent of
the power-law decreases with the growth of the threshold $s_{min}$.
Power-laws with negative exponents that are smaller in absolute
value of course decay slower. Thus, in this model,
correlations between events detected by using larger thresholds
are qualitatively different from the ones between small events, as they
have a slower decay with the waiting time.


Fig.~\ref{fig:M1_Ptw_resc} shows the same distributions rescaled by the
mean waiting time. It is evident that the distribution curves do
not collapse into the power-law with a given exponent, thus
representing the fact that the exponent values depend on the thresholds.

These results have a twofold meaning: they again show that
correlated avalanches can be found in SOC models. On the other hand, the
recent results on the data collapse of the waiting time distributions of
flares~\cite{flares} cannot be reproduced,
which means that the model is missing some important feature.
This should not be surprising, as our model is an extremely oversimplified
system.
It is also fair to say that the analysis of the results via
rescaling of the waiting time distributions is a test that has
not yet been  used to analyze data from models outside the SOC domain.

\subsection{Model II}

\begin{figure}[htb]
\includegraphics[angle=0,width=7.8cm]{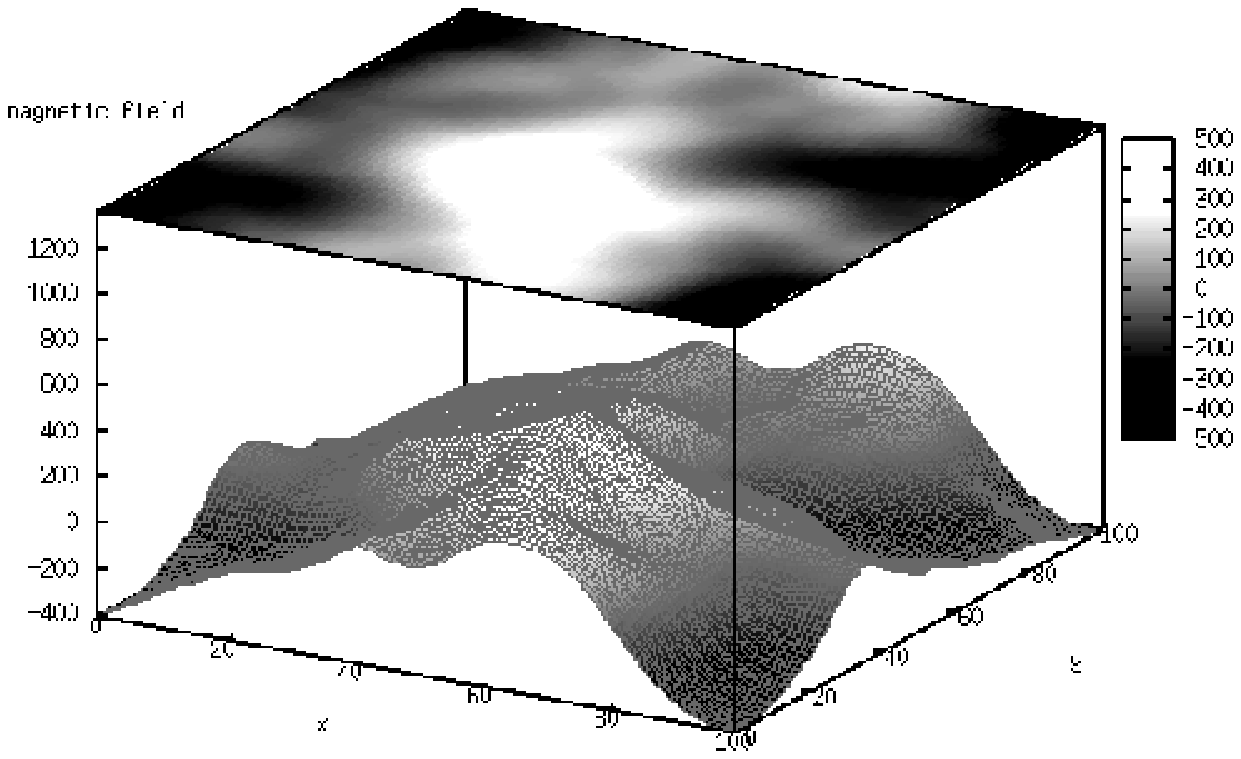}
\vskip 0.5truecm
\includegraphics[angle=0,width=7.8cm]{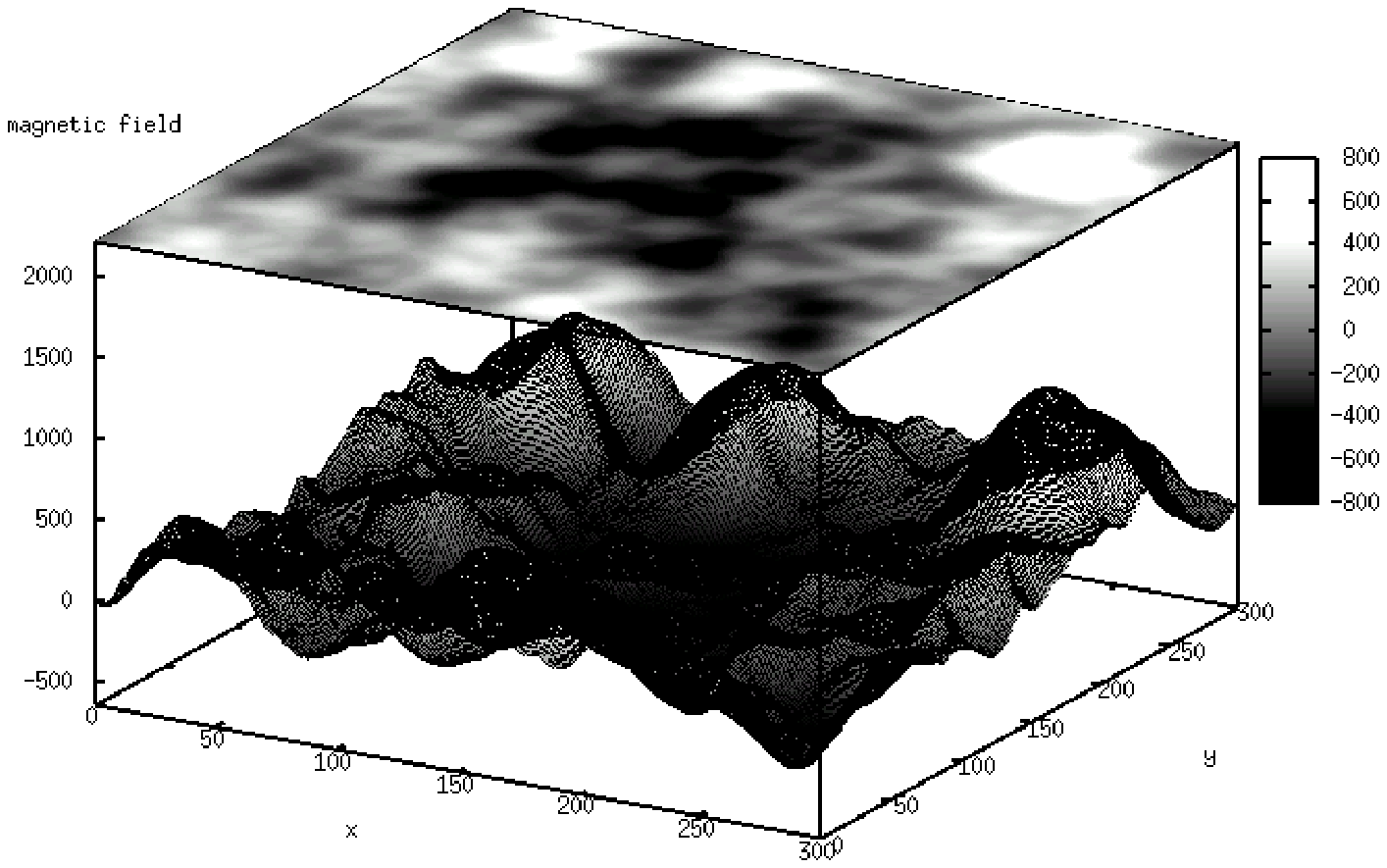}
\caption{Two configurations of model~II (with $p_{move}=10^{-2}$),
one for $L=100$ and one for $L=300$. \label{fig:M2_conf}}
\end{figure}

\begin{figure}[htb]
\includegraphics[angle=0,width=7.8cm]{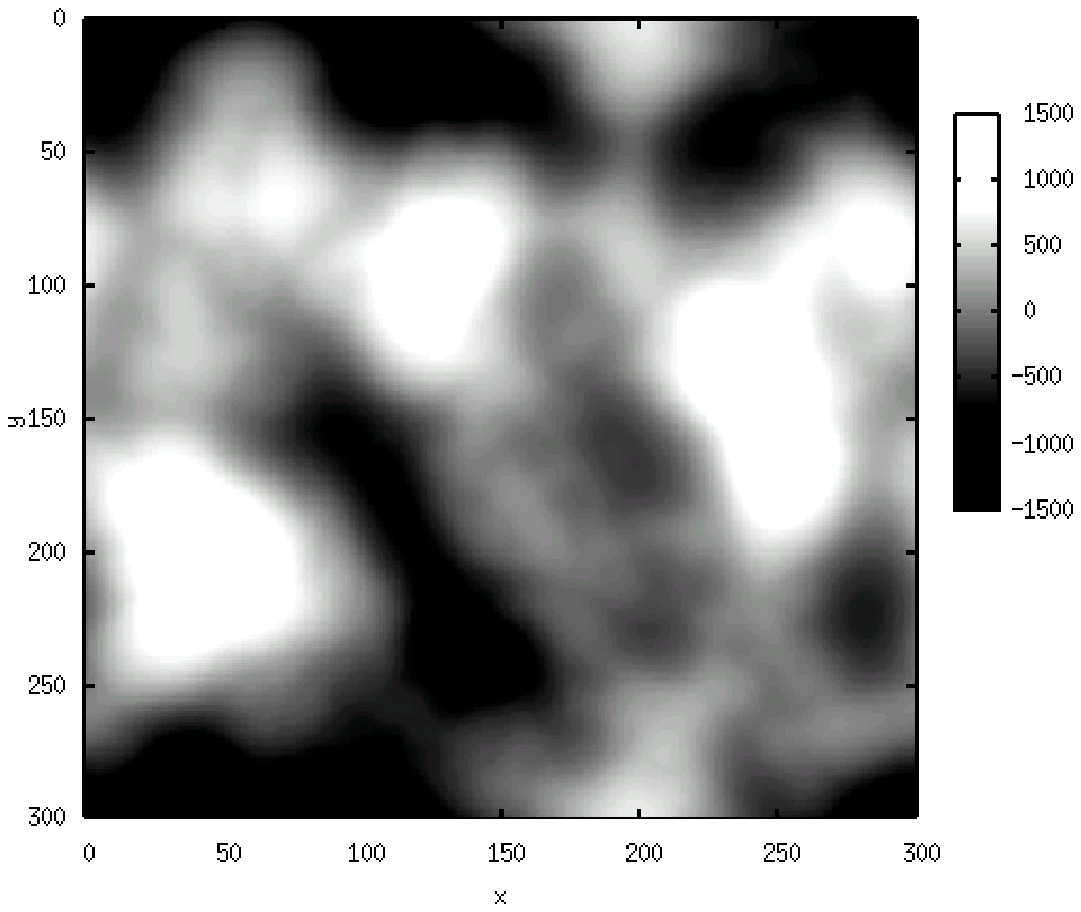}
\vskip 0.5truecm
\includegraphics[angle=0,width=7.8cm]{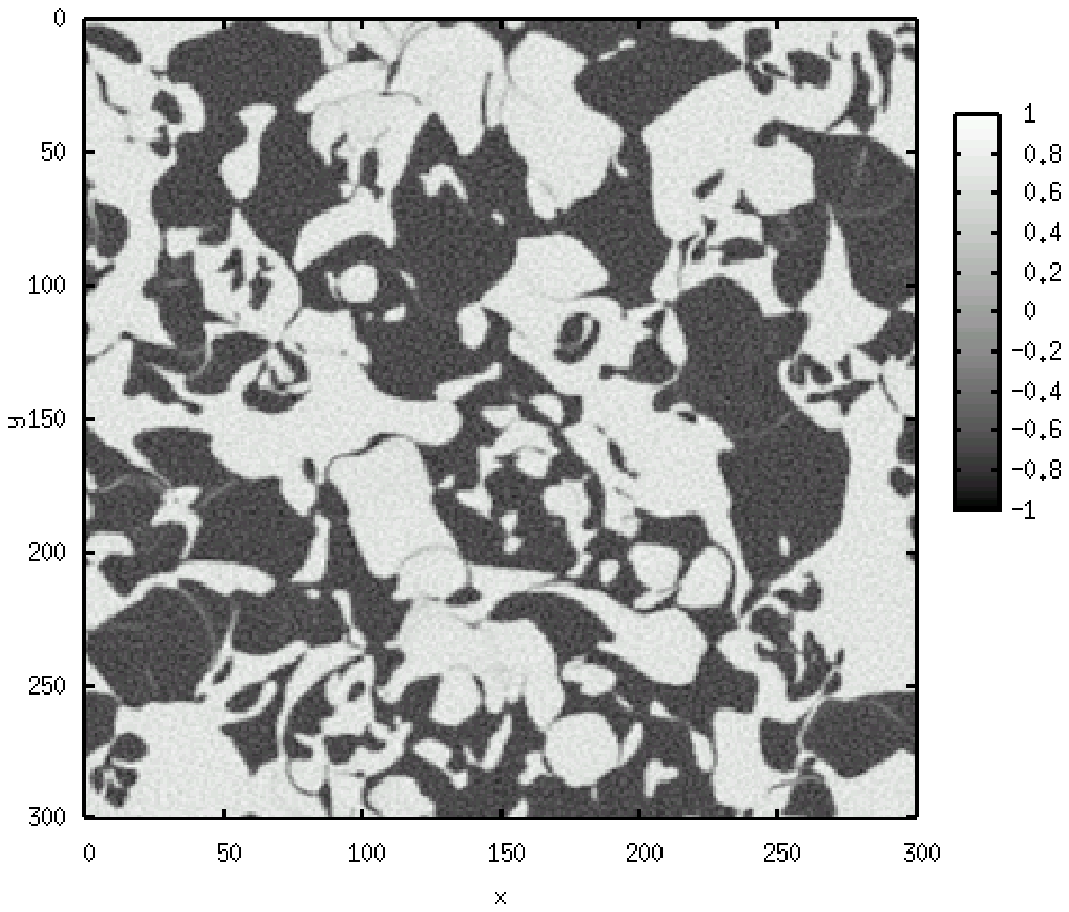}
\caption{Field $h$ (top panel) and curvature $\Delta h$ (bottom
panel) of a configuration for model II (with $L=300$ and
$p_{move}=10^{-2}$). \label{fig:M2_conf_B_dB} }
\end{figure}

Model II has two sources of opposite polarities that execute a
random walk as the result of an idealized complex convective
motion of an underlying turbulent atmosphere. While model~I
displays the usual profiles of the field $h$ (not shown), with a
maximum at the center of the lattice, we can observe far more
complex configurations in model~II. Two typical configurations of
the system, one for $L=100$ and one for $L=300$ (both with
$p_{move}=10^{-2}$), are shown in Fig.~\ref{fig:M2_conf}. One can
see that a non-trivial field landscape arises because of the
complex interplay between the relaxation dynamics and the
diffusion of the two sources of perturbation. In
Fig.~\ref{fig:M2_conf_B_dB} one can appreciate that there is also
a non-trivial relation between the field $h$ (top panel) and its
corresponding curvature field $\Delta h$ (bottom panel).



It is possible that the system dynamics generates a length scale
corresponding e.g.~to
the average distance between local minima and local maxima:
the comparison between the two configurations in Fig~\ref{fig:M2_conf},
$L=100$ and $L=300$, in this case seems to confirm this hypothesis, as the
typical size of the white and black areas in both ``magnetograms'' are similar.
This length scale contrasts with the scale-free nature of SOC, as the lattice
size $L$ does.

\begin{figure}[htb]
\includegraphics[angle=0,width=7.8cm]{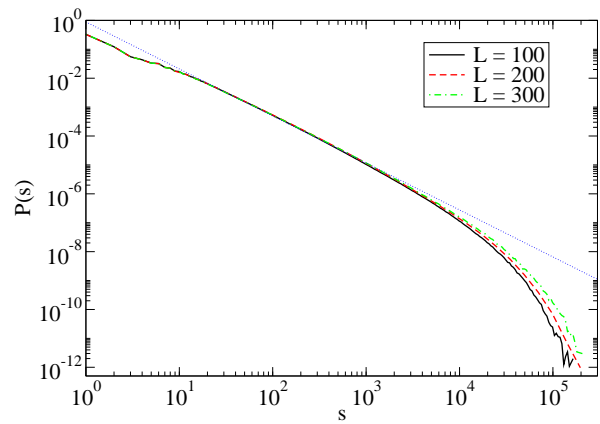}
\caption{Log-log plot of the distributions of avalanche sizes for model~II
(with $p_{move}=10^{-2}$) for lattice sizes $L=100$, $200$, and $300$.
The straight line interpolates the power-law range of the curves.
\label{fig:M2_Ps}}
\end{figure}

\begin{figure}[htb]
\includegraphics[angle=0,width=7.8cm]{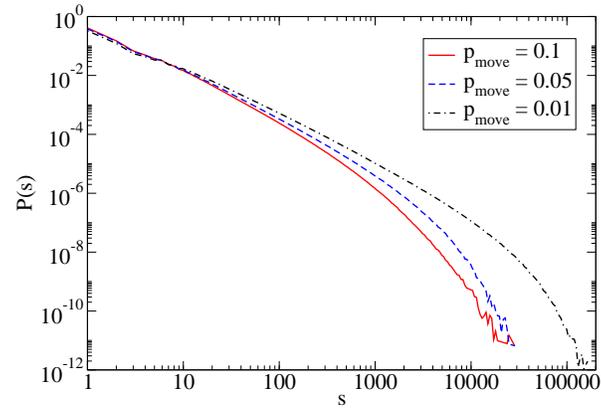}
\caption{Log-log plot of the distributions of avalanche sizes for model~II,
for $L=100$ and several $p_{move}$. \label{fig:M2_cut-Pmove}}
\end{figure}

A new length scale in the model, in addition to the lattice size, could
prevent the model from becoming asymptotically critical for $L\to \infty$.
This scenario is supported quantitatively by Fig.~\ref{fig:M2_Ps}, in which
the distribution of avalanche sizes, while displaying the usual power-law
range of SOC models, is cutoff at a size that essentially does not
scale with $L$. While for practical purposes this is not a problem
(we have a distribution with a power-law range that occupies some decades, like
experimental ones),
the mathematical assessment of criticality in SOC models would require also
this additional length scale to diverge. This might be achieved by letting
$p_{move}\to 0$, as suggested by figure \ref{fig:M2_cut-Pmove}, in which we see
that the size distribution develops a wider power-law range for
decreasing $p_{move}$. (Similar
results are found in several SOC models when some dissipation is introduced.
In our model the mobility of the sources might play a role similar
to the rate of dissipation during a toppling in  e.g.\ the Abelian sandpile.)
Unfortunately, the simulation of systems with a low $p_{move}$ is much more
time-consuming. Indeed, it becomes difficult to achieve a correct sampling,
because  it takes a too long time for the drivers to span a
significant fraction of the lattice sites. Thus, we cannot assess any
precise statement concerning this limit.

%

%

\begin{figure}[htb]
\includegraphics[angle=0,width=7.8cm]{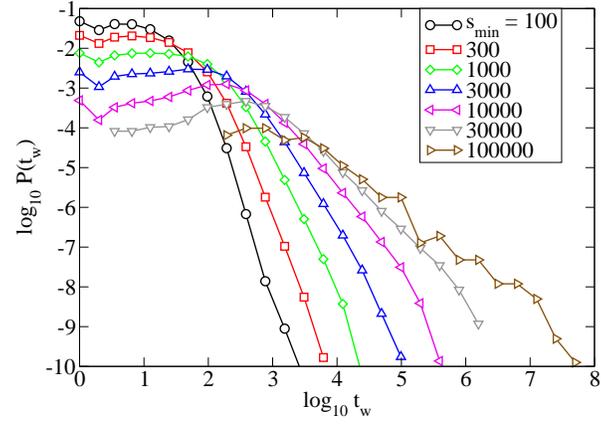}
\caption{Log-log plot of the distributions of waiting times for model~II
(with $p_{move}=10^{-2}$ and $L=200$) for several thresholds.
\label{fig:M2_tw}}
\end{figure}

Model II shows distribution of waiting times with the features that we
also found in model~I, see figure \ref{fig:M2_tw}. In particular, the
power-law tails of the distributions have different exponents and hence
the curves would not collapse upon rescaling of the times.

%

\section{conclusions}
The magnetized atmosphere of the Sun is represented in our model
as a magnetic "canopy". The magnetic field ideally is entering perpendicular
to a two-dimensional domain, with sources of perturbation
undergoing a diffusion and producing local instabilities
whenever the curvature of the magnetic field is too high.  These
instabilities are flattened out in a dynamics producing avalanches
of energy release that we interpret as flares.

In a first model, we have performed numerical simulations with
perturbation and boundary conditions typical of the LH-model,
and we have measured the size of the
avalanches and the time between avalanches of a size bigger than a
given threshold.  The analysis has shown that
the diffusive character of the feeding sources
is related to the power-law tail in the waiting time distributions between
avalanches.
The power-law exponents of these tails depend on the value of the
diffusion rate $p_{move}$ and on the thresholds in the size of the avalanches.
Thus, they are non-universal and the avalanche time series at a give scale
is quantitatively and also qualitatively different from the same time series
at another scale, at variance with real time series~\cite{flares}.

With two sources pumping perturbations of opposite polarities we obtain
a second model with novel features, like typical configurations with
a complex field landscape (magnetograms), and
with patches of positive and negative curvature that are non-trivially
related to the corresponding magnetic field.
When the sources move slow enough, also this system reaches a
critical regime with power-law distributions for
the avalanche sizes and waiting times.
This behavior is achieved even without a
mechanism of magnetic reconnection, and without open boundary conditions.
A length scale independent of the system size but depending
on the source diffusion rate is present in this case,
which prevents the system from approaching
a pure SOC state for larger and larger lattices.

In order to develop a more detailed model, which could be used for
a direct comparison with the observations, one should add more and
different types of mechanism of energy release, for example, like
the ones introduced in \cite{hughes2003,paczushug}. It would
ultimately amount to a systematic study of the processes of
transformation and redistribution of the magnetic energy.  The
results here already reveal that temporal correlations in the
energy accumulation and in the release processes can be expected
from a SOC model, realizing and incorporating the clustering in
space and time of the active areas of the system.

\vspace{5mm}
\noindent{\bf Acknowledgments}\\
Work of B.M.S. have been supported by short term post doctoral
grant at the Instituut voor Theoretische Fysica (ITF), K.U.Leuven, Grant
of K.U.Leuven -PDM/06/116 and Grant of Georgian National Science
Foundation - GNSF/ST06/4-098.
M.B.~thanks the ITF for the warm hospitality.
\bibliographystyle{plain}

\begin{thebibliography}{10}

\bibitem{BTW1987}
P. Bak, C. Tang, K. Wiesenfeld, Phys. Rev. Lett. 59 (1987) 381.

\bibitem{bak}
P. Bak, How Nature works: The Science of Self-Organized
Criticality, Copernicus, New York, 1996.
\bibitem{sornette}
D. Sornette, Critical Phenomena in Natural Sciences Springer,
Heidelberg, 2000.
\bibitem{jenbook}
H.J. Jensen, In Self-Organized Criticality, Cambridge University
Press, 1998.

\bibitem{aschwanden2}
M.J. Aschwanden, T.D. Tarbell, R.W. Nightingale, C.J. Schrijver,
A. Title, ApJ 535 (2000) 1047
\bibitem{aschwanden1}
M.J. Aschwanden, R.W. Nightingale, T.D. Tarbell, C.J. Wolfson, ApJ
535 (2000) 1027.


\bibitem{heyvaerts2001}
J. Heyvaerts, Coronal Heating Mechanisms, ed. P. Murdin, Bristol:
Institute of Physics Publishing 2001.
\bibitem{sherg1}
B.M. Shergelashvili, S. Poedts, A.D. Pataraya, ApJ 642 (2006) L73.
\bibitem{sherg2}
B.M. Shergelashvili, S. Poedts, A.D. Pataraya, Proceedings of the
11th European Solar Physics Meeting "The Dynamic Sun: Challenges
for Theory and Observations", (ESA SP-600), p.98 2005.

\bibitem{LH1991}
E.T. Lu, R.J. Hamilton, ApJ 380 (1991) L89
\bibitem{LU1993}
E.T. Lu, R.J. Hamilton, J.M. McTiernan, K.R. Bromund, ApJ 412
(1993) 841.
\bibitem{parker1989}
E.N. Parker, Sol. Phys. 121 (1989) 271.
\bibitem{sturrock1984}
P.A. Sturrock, P. Kaufman, R.L. Moore and D.F. Smith Sol. Phys. 94
(1984) 341.
\bibitem{sturrock1990}
P.A. Sturrock, W.W. Dixon, J.A. Klimchuk, S.K. Antiochos, ApJ 356
(1990) L31.
\bibitem{parker1988}
E.N. Parker, ApJ 330 (1988) 474.
\bibitem{isliker2000}
H. Isliker, A. Anastasiadis, L. Vhalos, A\&A 363 (2000) 1134.
\bibitem{vlahos2004}
L. Vlahos, H. Isliker, F. Lepreti, ApJ 608 (2004) 540.
\bibitem{krasnosels2002}
V. Krasnoselskikh, O. Podladchikova, B. Lefebvre, N. Vilmer A\&A
382 (2002) 699.
\bibitem{podlad2002}
O. Podladchikova, T. Dudok de Wit, V. Krasnoselskikh, B. Lefebvre,
A\&A 382 (2002) 713.
\bibitem{boffetta1999}
G. Boffetta, V. Carbone, P. Giuliani, P. Veltri, A. Vulpiani,
Phys. Rev. Lett. 83 (1999) 4662.
\bibitem{BM2006}
M. Baiesi, C. Maes, Europhys. Lett. 75 (2006) 413.
\bibitem{wood}
R. Woodard, D.E. Newman, R. S\'anchez, B.A. Carreras, Physica A
373 (2007) 215.
\bibitem{menech}
M. De Menech, A.L. Stella, Physica A 309 (2002) 289.
\bibitem{note_1}
There are other mechanisms for producing a nontrivial (critical)
temporal statistics (see~\cite{wood,BM2006,menech} and references
therein. For example, one can also confine the random driving to
the boundary of the domain, see: H.J.~Jensen,  Phys. Rev. Lett. 64
(1990) 3103.

\bibitem{earthquakes1}
\'A. Corral, Phys. Rev. E 68 (2003) 035102.
\bibitem{earthquakes2}
P. Bak, K. Christensen, L. Danon, T. Scanlon, Phys. Rev. Lett. 88
(2002) 178501.
\bibitem{flares}
M. Baiesi, M. Paczuski, A.L. Stella, Phys. Rev. Lett. 96 (2006)
051103.

\bibitem{PACZUSKI2005}
M. Paczuski, S. Boettcher, M. Baiesi, Phys. Rev. Lett. 95 (2005)
181102.

\bibitem{hughes2003}
D. Hughes, M. Paczuski, R.O. Dendy, P. Helander, K.G. McClements,
Phys. Rev. Lett. 90 (2003) 131101-1.
\bibitem{paczushug}
M. Paczuski, D. Hughes, Physica A. 342 (2004) 158.
\end{thebibliography}


\newpage

\newpage

\begin{figure}[htb]
\includegraphics[angle=0,width=7.8cm]{fig3.eps}
\caption{Log-log plot of the distributions of waiting-times for model~I
(with $L=150$ and $p_{move} = 10^{-1}$) for different values of the
threshold $s_{min}$. \label{fig:M1_Ptw_s} }%
\vskip 0.5 truecm
\includegraphics[angle=0,width=7.8cm]{fig4.eps}
\caption{As in Fig.~\ref{fig:M1_Ptw_s} for the distributions
rescaled by the mean waiting time. \label{fig:M1_Ptw_resc} }
\end{figure}

\end{document}